\documentclass[submission,copyright,creativecommons]{eptcs}
\providecommand{\event}{FMAS 2024} 

\usepackage{iftex}
\usepackage{graphicx}
\usepackage{todonotes}
\usepackage{subcaption}
\usepackage{wrapfig}

\usepackage{makecell}
\usepackage{tabularx}
\usepackage{booktabs}
\usepackage{multirow}
\usepackage{multicol}
\usepackage{longtable}
\usepackage{hhline}

\usepackage{siunitx}
\usepackage{amsmath}
\usepackage{cleveref}

\usepackage{tikz}
\usepackage{xcolor}
\usepackage{comment}

\definecolor{orcidlogocol}{HTML}{A6CE39}
\usetikzlibrary{svg.path}
\tikzset{
	orcidlogo/.pic={
		\fill[orcidlogocol] svg{M256,128c0,70.7-57.3,128-128,128C57.3,256,0,198.7,0,128C0,57.3,57.3,0,128,0C198.7,0,256,57.3,256,128z};
		\fill[white] svg{M86.3,186.2H70.9V79.1h15.4v48.4V186.2z}
		svg{M108.9,79.1h41.6c39.6,0,57,28.3,57,53.6c0,27.5-21.5,53.6-56.8,53.6h-41.8V79.1z M124.3,172.4h24.5c34.9,0,42.9-26.5,42.9-39.7c0-21.5-13.7-39.7-43.7-39.7h-23.7V172.4z}
		svg{M88.7,56.8c0,5.5-4.5,10.1-10.1,10.1c-5.6,0-10.1-4.6-10.1-10.1c0-5.6,4.5-10.1,10.1-10.1C84.2,46.7,88.7,51.3,88.7,56.8z};
	}
}
\newcommand{\orcidID}[1]{%
	\resizebox{8px}{8px}{
		\href{https://orcid.org/#1}{\tikz[yscale=-1,transform shape]{\pic{orcidlogo}}}}%
}

\newcommand{\prob}{{\sc ProB}}
\newcommand{\probtwoui}{\textsc{ProB2-UI}}

\newcommand{\simb}{\textsc{SimB}}

\ifpdf
  \usepackage{underscore}         
  \usepackage[T1]{fontenc}        
\else
  \usepackage{breakurl}           
\fi

\title{
Using Formal Models, Safety Shields and Certified Control to Validate AI-Based Train Systems
\thanks{This research is part of the KI-LOK project funded by the ``Bundesministerium f\"ur Wirtschaft und Energie''; grant \# 19/21007E.
The work of Fabian Vu is part of the IVOIRE project funded by ``Deutsche Forschungsgemeinschaft'' (DFG) and the Austrian Science Fund (FWF) grant \# I 4744-N.}}
\author{
	Jan Gruteser\orcidID{0009-0006-4228-404X}
    \quad\quad Jan Roßbach\orcidID{0009-0005-7725-9832}
    \quad\quad Fabian Vu\orcidID{0000-0003-2556-5553}
    \quad\quad Michael Leuschel \orcidID{0000-0002-4595-1518}
	\institute{Heinrich Heine University D\"{u}sseldorf\\
		Faculty of Mathematics and Natural Sciences\\
		Department of Computer Science\\
		D\"{u}sseldorf, Germany\\}
	\email{\quad \{jan.gruteser,jan.rossbach,fabian.vu,leuschel\}@hhu.de}
}
\def\titlerunning{Using Formal Models, Safety Shields and Certified Control to Validate AI-Based Train Systems}
\def\authorrunning{J. Gruteser, J. Roßbach, F. Vu \& M. Leuschel}
\begin{document}
\maketitle

%
%
%
%

\begin{abstract}
The certification of 
 autonomous systems is an important concern in science and industry.
The KI-LOK project explores new methods for certifying and safely integrating AI components into autonomous trains.
We pursued a two-layered approach: (1) ensuring the safety of the steering system by formal analysis using the B method,
and (2) improving the reliability of the perception system with a runtime certificate checker.
This work links both strategies within a demonstrator that runs simulations on the formal model, controlled by the real AI output and the real certificate checker.
The demonstrator is integrated into the validation tool \prob{}.
This enables runtime monitoring, runtime verification, and statistical validation of formal safety properties using a formal B model.
Consequently, one can detect and analyse potential vulnerabilities and weaknesses of the AI and the certificate checker.
We apply these techniques to a signal detection case study and present our findings. 

\end{abstract}

\section{Introduction and Motivation}

Artificial intelligence (AI) is increasingly used in safety-critical applications such as autonomous driving~\cite{sun2020scalability} and autonomous flying~\cite{nonami2010autonomous,taxinet}.
While AI can be effective for many challenging tasks, it also introduces new risks and concerns.
This leads to new challenges regarding certification and ensuring the safety of AI components
 (see, e.g., Peleska et al.~\cite{peleska2022standardisation} 
  in the context of  autonomous railway systems).

This work deals with systems that employ an AI perception system, such as image recognition.
Those systems include autonomous vehicles and autonomous railway systems.
In earlier work~\cite{gruteser2023formal}, we formally verified a steering system, assuming the perception system works perfectly.
However, as the perception system is imperfect, we also created simulations with (hand-coded) probabilities for all kinds of erroneous detections.
We then applied Monte Carlo simulation to estimate the likelihood of safety-critical errors.
As a concrete case study, we applied those techniques to an AI-based railway system~\cite{gruteser2023formal}
using \prob{}~\cite{probjournal} and \simb{}~\cite{simb}.
In this paper, we move towards using the {\em real\/} AI within these simulation and validation runs, rather than using estimated error rates.
As outlined by Myllyaho et al.~\cite{literatureValidationAiSystems}, fully virtual simulation enables validation of the system in dangerous situations
without real danger.
Although the validity of the simulator is difficult to verify, simulation still helps as a validation method
to initially assess the quality of the system under evaluation.

The AI perception system itself is based on the widely-used
YOLO~\cite{DBLP:journals/corr/RedmonDGF15} architecture,
making the system difficult to verify with formal methods alone.
To address this, we implemented a runtime certificate checker
using classical computer vision algorithms to verify the output of the AI~\cite{Rossbach2023}.
This checker can be certified using classical techniques (e.g., \cite{EN50128}).

This work presents a real-time demonstrator linking the formal model, the AI, and certified control, extending an approach used to validate reinforcement learning agents~\cite{vu2024validation}.
  AI now controls the simulation directly by executing events for the perception system in the formal model.
With our approach, we capture real AI behaviour and can use
  two runtime monitoring techniques:
  1) the formal B model acts as a safety shield for the steering system (e.g., to detect false negatives),
  2) and the certified control monitors the perception system (detecting false positives).
We demonstrate this methodology on a railway case study~\cite{gruteser2023formal},
and discuss our findings and the challenges of the approach.


\paragraph{B Method, ProB, and SimB.}

The B method~\cite{bmethod} is a state-based formal method for specifying and verifying software systems.
The B method is based on first-order logic and set theory and has been used industrially for over 25 years \cite{DBLP:conf/fmics/ButlerKKLLMV20}
 to generate software that is ``correct by construction'' \cite{DBLP:journals/tsi/DolleEF03,Siemens:B2007}, and for system-level safety modelling.
For the latter, the B method has been used for many railway applications, such as
  ETCS Hybrid Level 3~\cite{DBLP:journals/sttt/HansenLKKNNSS20,etcs-proof}
  and CBTC systems~\cite{DBLP:conf/rssrail/ComptierDPMTS17,DBLP:conf/rssrail/ComptierLMPM19}.
In this article, we use the B method to model autonomous train control in a shunting yard, based on the model from \cite{gruteser2023formal}.

\prob{}~\cite{probjournal} is an animator, constraint solver, and model checker for formal models.
It supports various formal languages including the formal B method.
\simb{}~\cite{simb} is a simulator built on \prob{}'s animator, supporting real-time simulation and Monte Carlo simulation. 
\simb{} can be linked with external software components~\cite{vu2024validation}. 
We use \prob{} and \simb{} in this article to run the steering system and the safety shield, and also for validating the entire system.

%
%
%

\paragraph{Certified Control.}

\begin{wrapfigure}{r}{0.25\textwidth}
	\vspace{-1em}
	\centering
	\begin{subfigure}{0.20\textwidth}
		\centering
		\includegraphics[width=\linewidth]{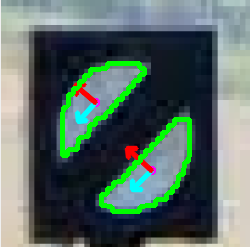}
		\caption{\centering Validated Sh1 Sign}
		\label{fig:checking-b}
	\end{subfigure}
	\vfill
	\begin{subfigure}{0.20\textwidth}
		\centering
		\includegraphics[width=\linewidth]{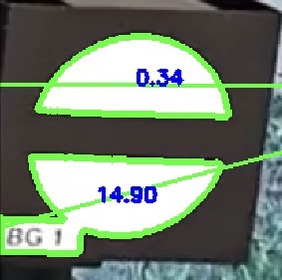}
		\caption{\centering Falsely rejected Sh0 Sign}
	    \label{fig:checking-c}
	\end{subfigure}
	\caption{\centering Runtime Monitor Example}
	\label{fig:certctrl}
	\vspace{-1em}
\end{wrapfigure}
Certified control~\cite{certcontrol} is a runtime monitoring approach to
ensure the safety of the perception system in autonomous vehicles.
Unlike conventional monitoring methods, certified control does not rely on independent perceptions.
Instead, a controller provides a \emph{certificate} containing all essential information to prove formal properties.
This certificate may be generated by a sophisticated AI algorithm, which does not need to be formally verified.
Using this certificate, the runtime monitor verifies if the specified criteria hold for the provided data.
This monitor can, in contrast to the AI system, be comparatively small and deterministic.
The architecture establishes a trusted foundation that can potentially be subjected
to a rigorous formal verification process.

In previous tests, this technique almost eliminated all false
positive detections, in exchange for rejecting some true positives~\cite{Rossbach2023}.
The bounding box of the sign detected by the YOLO model is cropped from the image and
passed to the runtime monitor, which uses various computer vision
techniques, e.g. contour detection, to validate the sign for expected features.
If the desired features are not recognised, the detection is rejected.
Figure~\ref{fig:certctrl} shows a successfully validated Sh1, and a Sh0 sign that would be rejected by the checker.

\section{Linking Formal Model, AI, and Certified Control}
\label{sec:ai-sim}

\begin{figure}[t]
	\begin{center}
		\includegraphics[width=0.9\linewidth]{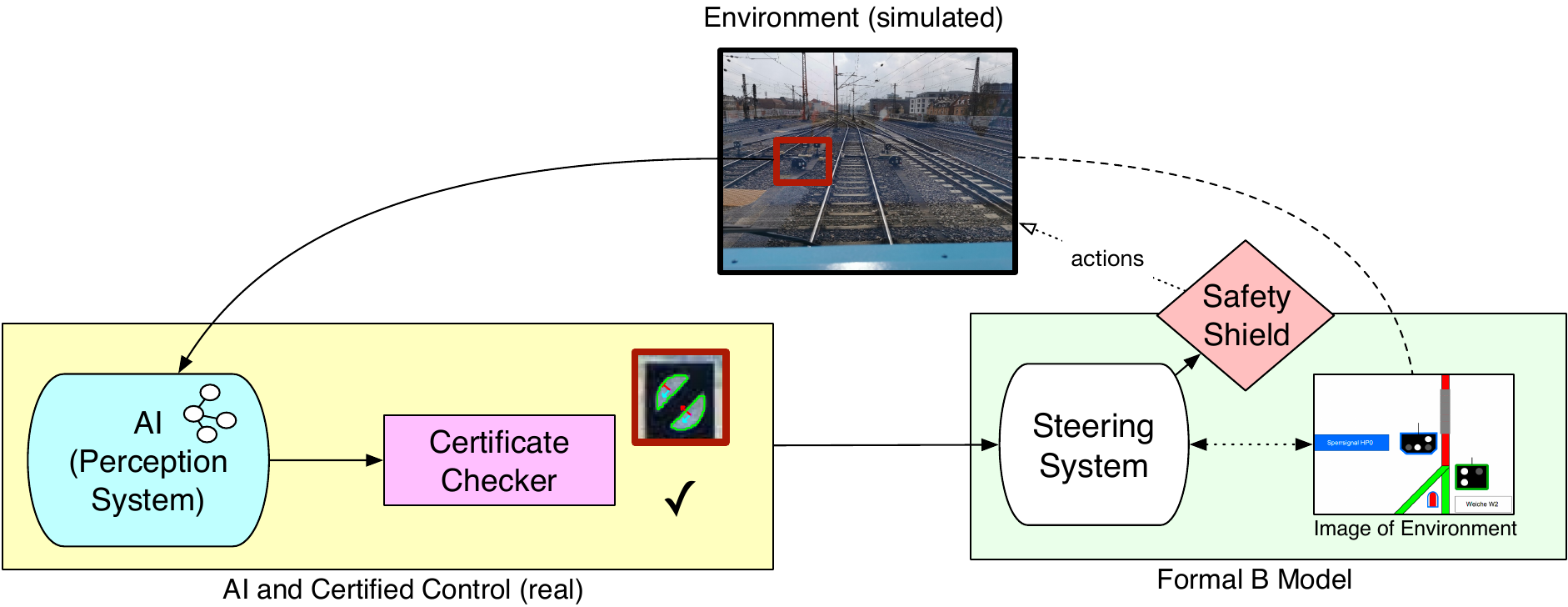}
	\end{center}
	\vspace{-1em}
	\caption{Simulation of the Formal B Model with AI-based Perception System and Certificate Checker in an Environment}
	\vspace{-0.5em}
	\label{fig-overview}
\end{figure}

This section describes how we link together the formal model, the AI, and the certificate checker.
In the formal model, we formally specify and verify the steering system
which includes safety shields to prevent unsafe operations based on the AI's perception
and the known environment.
In previous work~\cite{gruteser2023formal} on validating an AI-based train control system,
 we encoded probabilities for false positive and false negative detections of the AI by hand.
Now, we use {\em real\/} AI components and real certificate checkers for simulation.
Figure~\ref{fig-overview} gives an overview of how we link the formal model, the AI, and certified control inside a real-time demonstrator with runtime monitoring/verification.

In our case study, the AI-based perception system processes the environment in form of images at runtime.
Ideally, various techniques should be employed to ensure that the AI is trained correctly and performs well
  (see \cite{Rossbach2024} and references therein).
Additionally, we use certified control to monitor the perception system, and detect false positives
 (i.e., the monitor will reject detections which it cannot confirm).
The output of the perception system and the certificate checker are then synchronised with the formal model.
The simulated environment, including the image provided to the AI, must correspond to the formal model's current state.
This is a significant challenge as discussed in Section~\ref{sec:challenges}.
The formal model contains events for both the steering system and the perception system.
The model should be safe under the assumption that the perception system works perfectly.
Furthermore, the formal model can be used as a basis for a safety shield which enforces safe actions on the steering system.
The shield can disallow unsafe actions, like driving through a detected stop signal.
The safety shield can also detect false negatives.
For example, when the AI detects no signals but the formal model ``knows'' that a signal must be visible at the
 current location, it can enforce a safe fallback action (like stopping).

\section{Case Study: AI-based Signal Detection}
We apply the presented technique to a case study provided by our project partners (see~\cite{gruteser2023formal}). 
For this case study, we developed a formal B model~\cite{gruteser2023formal}, consisting of an environment, the steering system, and the perception system.
The environment includes obstacles, points (aka switches) and signal states, field elements and movements of the steered train.
The formal model abstracts away the AI-based perception system by events that represent possible outcomes
of the object detection, including correct, false positive and false negative detections.

The objective, or the ``mission order'', is to drive autonomously from the starting position through a small shunting yard to the destination without dangerous situations or at least as safely as human drivers~(cf. \cite[\textbf{PROB1-2}]{gruteser2023formal}). 
The focus of this work is the detection of signal aspects during the shunting movement.



\subsection{Implementation}

We collected images from multiple videos of the case study track containing the various signal aspects, e.g., stop signals, permission signals, and no signals along the route to capture an interactively changing environment.
Based on the position of the train and the state controlled by the formal model, an image with the appropriate signalling aspect is randomly selected from the corresponding collection.
For our experiments, we assume that a signal becomes visible as soon as the train is closer than 10 distance units (freely selectable).
The procedure for a simulation step is as follows:

\begin{enumerate}
	\item Pick an image randomly depending on the current train location and signal states in the B model
	\item Pass the image to our fine-tuned YOLOv8 model (cf.~\cite{Rossbach2023}); based on the result: if no signal has been detected: ignore and do not execute any operation in the B model; if a signal has been detected:
	\item[] \begin{itemize}
		    \item Correct signal (corresponding B event is enabled): execute \texttt{VIS\_DetectCorrectSignal}
	    	\item Wrong stop signal: execute \texttt{VIS\_DetectWrongStopSignal}
	    	\item Wrong permission signal: execute \texttt{VIS\_DetectWrongPermissionSignal}
    	\end{itemize}
	\item Execute event for environment change, e.g., switch signals or activate derailer, with a probability of 25\% or move train forwards and update controller with a probability of 75\% (environment changes should occur less frequently than train movements)
\end{enumerate}
The simulation runs in a loop until reaching the ending condition which is later explained in our experiments.

The controller of the steering system is updated after each detection to recompute the maximum allowed movement distance.
Since the simple object detection AI does not provide positioning information, we place all detections in the formal model at a fixed distance in front of the train.
In the second step, we (optionally) apply the certificate checker which monitors the output of the AI by accepting or rejecting its detections.

For the AI to run the simulation, we use \simb{}'s interface for external simulation~\cite{vu2024validation}.

\subsection{Experiments and First Results}

\begin{figure}[ht]
	\begin{center}
		\includegraphics[width=0.8\linewidth]{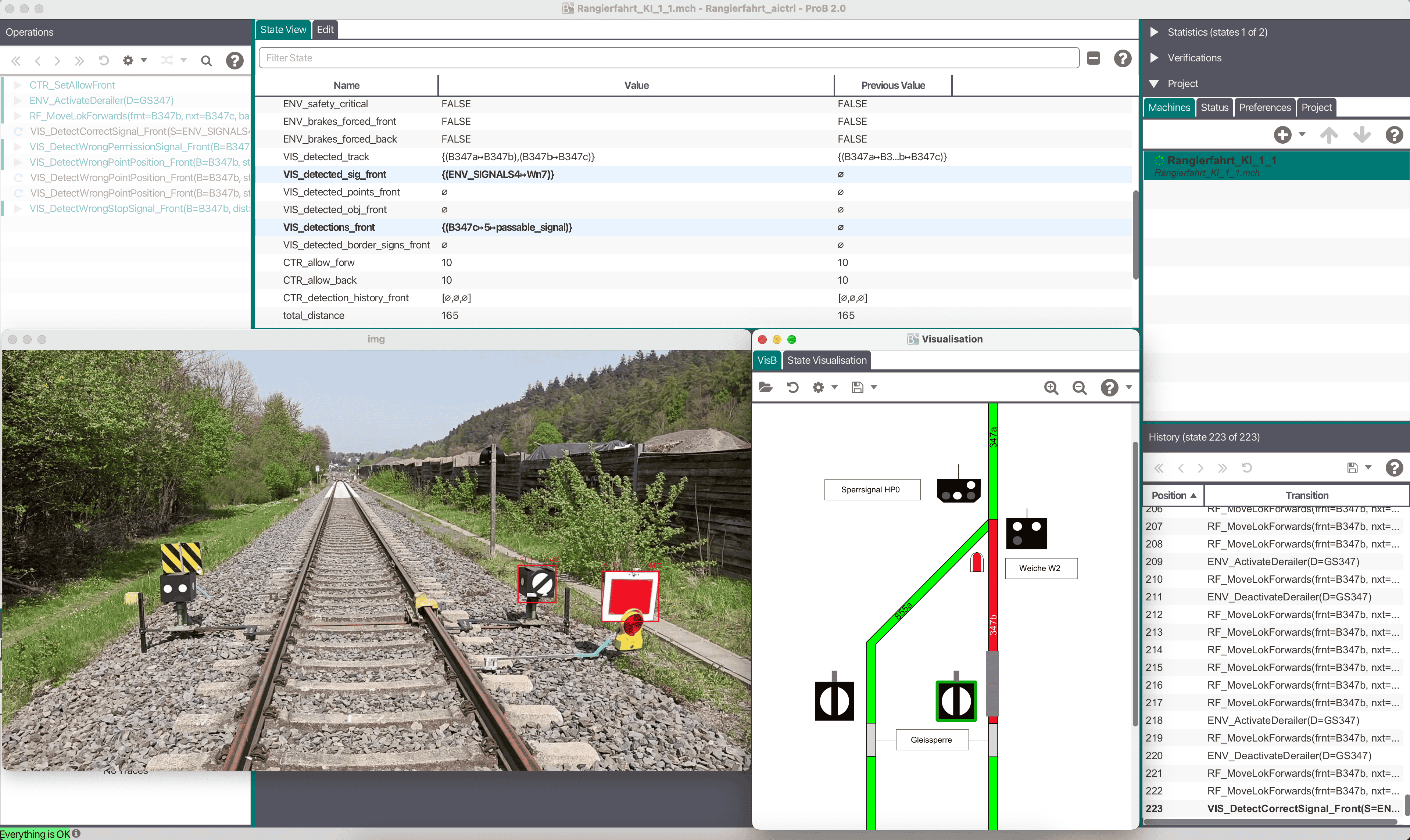}
	\end{center}
	\caption{Real-time Simulation in ProB2-UI together with Domain-Specific Visualisation and corresponding image; Signal is detected correctly.}
	\label{fig-prob2-ui}
\end{figure}

For initial experiments, we encoded a safety shield in the B model that only allows for signal detections at known positions of signals.
If no signal is detected at an expected position, the B~operations for train movements are disabled so that the train falls back to safe mode and stops in front of the signal.
This assumes that we have a map of the shunting yard and know at which locations the signals are located.
We then analyse the behaviour using \simb{}'s real-time simulation which is now controlled by the AI and the certificate checker.
For better understanding, we use the domain-specific VisB visualisation~\cite{visb} from previous work~\cite{gruteser2023formal}.
Both tools are part of \probtwoui{}~\cite{prob2ui}; an illustration is shown in Figure~\ref{fig-prob2-ui}.

With \simb{}, we also run Monte Carlo simulation with 500 runs for all combinations with/without safety shield, and with/without certified control.
We also investigated the effect of not applying the certificate checker to stop signals, so that these cannot be falsely rejected and the train always enters a safe state (stop).
The termination condition has been defined so that a simulation stops when a safety-critical situation occurs or when the train can no longer proceed, either due to its arrival at a stop signal or reaching the end.
For each execution run, we estimate the maximum (safe) \textit{distance} travelled to validate that the train does move forward (not driving at all would be 100 \% safe, but not useful).
Furthermore, we estimate the likelihood of a safety-critical situation where an accident might occur, similar to~\cite{gruteser2023formal}, using the safety properties \textbf{SAF1-5}
\footnote{Note that compared to the results in~\cite{gruteser2023formal}, we simulate \emph{real} AI behaviour at runtime, instead of encoding fixed probabilities for how the AI, i.e., the perception system could behave.}.
The results are shown in Table~\ref{tab:sim-results}.

\begin{table}[ht]
	\centering
	\caption{Results after 500 Simulations (NoStop: certificate checker is not applied to stop signals, False/Correct Det.: total number of activated operations for false/correct detections in all simulations)}
	\vspace{0.5em}
	\begin{tabular}{|c||c|c|c|c|c|c|}
		\cline{2-7}
		\multicolumn{1}{c|}{Controller} & \multicolumn{3}{c|}{No safety shield} & \multicolumn{3}{c|}{Safety Shield: Known signal positions} \\
		\multicolumn{1}{c|}{Cert. Control}  & No & NoStop & Yes & No & NoStop & Yes \\
		\cline{2-7}
		\hhline{-:======}%
		Distance & \num{6.4} \scriptsize{(0.0 \%)} & \num{6.1} \scriptsize{(0.0 \%)} & \num{221.2} \scriptsize{(63.2 \%)} & \num{247.8} \scriptsize{(82.4 \%)} & \num{247.0} \scriptsize{(80.4 \%)} & \num{216.8} \scriptsize{(63.0 \%)} \\
		Safe & 100 \% & 100 \% & 63.2 \% & 82.8 \% & 80.4 \% & 63.0 \% \\
		\cline{2-7}
		False Det. & \num{500} & \num{500} & \num{0} & \num{20611} & \num{20993} & \num{0} \\
		Correct Det. & - & - & \num{2335} & \num{9155} & \num{8929} & \num{2094} \\
		\hline
	\end{tabular}
	\label{tab:sim-results}
	\vspace{-1em}
\end{table}%

With certified control, it can be obtained that \textit{all} false detections are correctly rejected (in our simple environment).
Unfortunately, we also observe a significant decrease in correct detections with certified control due to false rejections (as shown in the last row of Table~\ref{tab:sim-results}).
When using the safety shield, there are still many false detections, as all detections are forwarded to the controller but then ignored if they were not expected.
Without safety shield and without certified control it can be observed that the train always stops early before the first wrong detected stop signal and
therefore does not make progress (the train does not even arrive at a position where it could detect a signal correctly).
This is because our detection model produces quite a lot of false positive detections of stop signals.
As expected, these \textit{false positive} detections can be effectively avoided by both the safety shield and the certificate checker.
This is reflected in significantly improved values for the distance travelled in the first row of Table~\ref{tab:sim-results}.

Surprisingly, the results of the combination of the safety shield and certified control are worse than those without certified control.
With \prob{}, we identify the cause by inspecting the execution runs simulated in \simb{}.
We found that \textit{falsely rejected} detections by the certificate checker can still lead to safety-critical situations when a signal has been correctly detected as a permission signal but then falls back to stop.
In this case, the safety shield no longer applies and a false negative detection can still cause the train to overrun the signal.
Moreover, we identified particular scenarios, in which the checker encounters difficulties in certifying a correct stop signal detection, resulting in unexpectedly many false rejections~(e.g.~Fig.~\ref{fig:checking-c}).

In general, it can be observed that the likelihood of safety-critical situations is still high.
This might be the case because of two reasons: (1) our custom AI model does not produce perfect results, and (2) there are still false negative detections.
Our experiments revealed that the safety shield of known signal positions is still not enough to tackle these.



\subsection{Challenges}\label{sec:challenges}

The main benefit of our methodology is that we can link the execution of a real AI model to a formal model
and check its behaviour using formal properties and statistical validation techniques.
However, there are still many challenges and limitations.

A major issue is to match the real environment provided to the AI with the environment's state in the formal model.
This requires an interactive simulation environment with control of the environment (signals, points, etc.) and control of the train,
to ensure that the actuators controlled by the formal model are taken into account by the simulated environment.
We have conducted some successful experiments with a commercial train simulator, but since it is not designed for discrete states, as required by the formal model, the handling is complicated (apart from the fact that a new instance of the simulator would have to be set up for each simulation run).
Simply using a video as input is not sufficient either, as the environment remains static and we have no control over the movement of the train.
In our current experiments, we avoid this problem by sampling images from videos fitting to the current state.
Although this is sufficient to demonstrate the concept, we have not yet simulated real train rides.
This requires a simulation environment for configurable scenarios (with \prob{} we are already able to load and visualise flexible scenarios, e.g., via a standardised data exchange format such as railML~\cite{gruteser-railml}).
While there are already established tools in the automotive sector, e.g., CARLA~\cite{carla}, such tools are still rare in the railway sector, but are under development~\cite{trainsimdataset,fraunhoferSimEnv,wild2023towards}.

\section{Related Work}

There are many approaches to verifying neural networks~\cite{katz2017reluplex,ruan2018reachability,verification-dnn}.
In practical application, however, it is challenging for these techniques to scale to large neural networks.
Another technique is robustness checks~\cite{gehr2018ai2,robustness} which also work on neural networks.
Robustness checks aim to ensure the safety of the AI directly, while this work employs safety boxes around the AI.
For instance, the perception system could be unsafe, but is monitored by a certificate checker.
Similarly, the steering system could make unsafe decisions based on the perception, but is monitored by a safety shield encoded in the formal model.

Another work presented by Pasareanu et al.~\cite{perception-DNN} abstracts away perception components,
and replaces them with a probabilistic component that estimates their behaviour.
In particular, the probabilities are derived from confusion matrices computed for the underlying neural network.
Finally, the verification is done by probabilistic model checkers such as PRISM~\cite{kwiatkowska2011prism} and STORM~\cite{hensel2022probabilistic}.
To improve safety, Pasareanu et al.~\cite{perception-DNN} employ run-time guards which are used as runtime monitors.

Instead of estimating the probabilities for real AI behaviour, this work simulates real AI behaviour at runtime.
This means that the perception system operates at runtime with real images and provides the detection to the validation tools \prob{} and \simb{}.
Alternatively, we could have extracted confusion matrices and encoded them as probabilities into the simulation.

\section{Conclusion and Outlook}
This work successfully links a formal model, AI, and certified control to a real-time demonstrator.
We demonstrated the technique in an AI-based train system.
The methodology consists of the following steps:
(1) formally specify and verify the steering system using formal methods,
(2) encode safety shields in the formal model to prevent unsafe operations,
(3) use \emph{real} AI for simulation,
(4) add a runtime certificate checker of the AI outputs to reduce false positive detections,
and (5) link all components to simulate the formal model with the AI and the certificate checker's output.
Using the tools \prob{} and \simb{}, we then evaluate the performance of the AI with (and without) certified control and safety shield.

With \simb{}, we can make statistical statements about the formal properties of the whole system.
We analysed the likelihood of unsafe situations and identified weaknesses in our AI and certificate checker.
With our approach, we can identify issues early during development.
The results are then used to improve the AI or the certificate checker, followed by further validation.

In our case study, we identified false negatives of certified control as a cause of unsafe situations.
In future, we thus need to improve our safety shield, to better protect against such false negative detections and reduce the error rates to levels required for certification.
Other future improvements are to link our tool with a realistic simulation environment, e.g., in the form of co-simulation.  

\paragraph{Acknowledgements.}
We thank Hitachi for providing the case study and parts of the video material.

\bibliographystyle{eptcs}
\bibliography{references}

\begin{thebibliography}{10}
\providecommand{\bibitemdeclare}[2]{}
\providecommand{\surnamestart}{}
\providecommand{\surnameend}{}
\providecommand{\urlprefix}{Available at }
\providecommand{\url}[1]{\texttt{#1}}
\providecommand{\href}[2]{\texttt{#2}}
\providecommand{\urlalt}[2]{\href{#1}{#2}}
\providecommand{\doi}[1]{doi:\urlalt{https://doi.org/#1}{#1}}
\providecommand{\eprint}[1]{arXiv:\urlalt{https://arxiv.org/abs/#1}{#1}}
\providecommand{\bibinfo}[2]{#2}

\bibitemdeclare{book}{bmethod}
\bibitem{bmethod}
\bibinfo{author}{Jean-Raymond \surnamestart Abrial\surnameend} \&
  \bibinfo{author}{A.~\surnamestart Hoare\surnameend} (\bibinfo{year}{2005}):
  \emph{\bibinfo{title}{{The B-Book: Assigning Programs to Meanings}}}.
\newblock \bibinfo{publisher}{Cambridge University Press},
  \doi{10.1017/CBO9780511624162}.

\bibitemdeclare{inproceedings}{prob2ui}
\bibitem{prob2ui}
\bibinfo{author}{Jens \surnamestart Bendisposto\surnameend},
  \bibinfo{author}{David \surnamestart Gele{\ss}us\surnameend},
  \bibinfo{author}{Yumiko \surnamestart Jansing\surnameend},
  \bibinfo{author}{Michael \surnamestart Leuschel\surnameend},
  \bibinfo{author}{Antonia \surnamestart P{\"u}tz\surnameend},
  \bibinfo{author}{Fabian \surnamestart Vu\surnameend} \&
  \bibinfo{author}{Michelle \surnamestart Werth\surnameend}
  (\bibinfo{year}{2021}): \emph{\bibinfo{title}{ProB2-UI: A Java-Based User
  Interface for ProB}}.
\newblock In: {\slshape \bibinfo{booktitle}{Proceedings FMICS}}, {\slshape
  \bibinfo{series}{LNCS}} \bibinfo{volume}{12863},
  \bibinfo{organization}{Springer}, pp. \bibinfo{pages}{193--201},
  \doi{10.1007/978-3-030-85248-1_12}.

\bibitemdeclare{inproceedings}{DBLP:conf/fmics/ButlerKKLLMV20}
\bibitem{DBLP:conf/fmics/ButlerKKLLMV20}
\bibinfo{author}{Michael~J. \surnamestart Butler\surnameend},
  \bibinfo{author}{Philipp \surnamestart K{\"{o}}rner\surnameend},
  \bibinfo{author}{Sebastian \surnamestart Krings\surnameend},
  \bibinfo{author}{Thierry \surnamestart Lecomte\surnameend},
  \bibinfo{author}{Michael \surnamestart Leuschel\surnameend},
  \bibinfo{author}{Luis{-}Fernando \surnamestart Mejia\surnameend} \&
  \bibinfo{author}{Laurent \surnamestart Voisin\surnameend}
  (\bibinfo{year}{2020}): \emph{\bibinfo{title}{The First Twenty-Five Years of
  Industrial Use of the {B}-Method}}.
\newblock In: {\slshape \bibinfo{booktitle}{Proceedings {FMICS}}},
  \bibinfo{series}{LNCS 12327}, pp. \bibinfo{pages}{189--209},
  \doi{10.1007/978-3-030-58298-2\_8}.

\bibitemdeclare{techreport}{EN50128}
\bibitem{EN50128}
\bibinfo{author}{\surnamestart CENELEC\surnameend} (\bibinfo{year}{2011}):
  \emph{\bibinfo{title}{Railway Applications -- Communication, signalling and
  processing systems -- Software for railway control and protection systems}}.
\newblock \bibinfo{type}{Technical Report} \bibinfo{number}{EN50128},
  \bibinfo{institution}{European Standard}.

\bibitemdeclare{inproceedings}{DBLP:conf/rssrail/ComptierDPMTS17}
\bibitem{DBLP:conf/rssrail/ComptierDPMTS17}
\bibinfo{author}{Mathieu \surnamestart Comptier\surnameend},
  \bibinfo{author}{David \surnamestart D{\'{e}}harbe\surnameend},
  \bibinfo{author}{Julien~Molinero \surnamestart Perez\surnameend},
  \bibinfo{author}{Louis \surnamestart Mussat\surnameend},
  \bibinfo{author}{Pierre \surnamestart Thibaut\surnameend} \&
  \bibinfo{author}{Denis \surnamestart Sabatier\surnameend}
  (\bibinfo{year}{2017}): \emph{\bibinfo{title}{Safety Analysis of a {CBTC}
  System: {A} Rigorous Approach with {Event-B}}}.
\newblock In: {\slshape \bibinfo{booktitle}{Proceedings RSSRail}}, pp.
  \bibinfo{pages}{148--159}, \doi{10.1007/978-3-319-68499-4\_10}.

\bibitemdeclare{inproceedings}{DBLP:conf/rssrail/ComptierLMPM19}
\bibitem{DBLP:conf/rssrail/ComptierLMPM19}
\bibinfo{author}{Mathieu \surnamestart Comptier\surnameend},
  \bibinfo{author}{Michael \surnamestart Leuschel\surnameend},
  \bibinfo{author}{Luis{-}Fernando \surnamestart Mejia\surnameend},
  \bibinfo{author}{Julien~Molinero \surnamestart Perez\surnameend} \&
  \bibinfo{author}{Mareike \surnamestart Mutz\surnameend}
  (\bibinfo{year}{2019}): \emph{\bibinfo{title}{Property-Based Modelling and
  Validation of a {CBTC} Zone Controller in {Event-B}}}.
\newblock In: {\slshape \bibinfo{booktitle}{Proceedings RSSRail}}, pp.
  \bibinfo{pages}{202--212}, \doi{10.1007/978-3-030-18744-6\_13}.

\bibitemdeclare{article}{DBLP:journals/tsi/DolleEF03}
\bibitem{DBLP:journals/tsi/DolleEF03}
\bibinfo{author}{Daniel \surnamestart Doll{\'e}\surnameend},
  \bibinfo{author}{Didier \surnamestart Essam{\'e}\surnameend} \&
  \bibinfo{author}{J{\'e}r{\^o}me \surnamestart Falampin\surnameend}
  (\bibinfo{year}{2003}): \emph{\bibinfo{title}{B dans le transport
  ferroviaire. {L}'exp{\'e}rience de {Siemens} {T}ransportation {S}ystems.}}
\newblock {\slshape \bibinfo{journal}{Technique et Science Informatiques}}
  \bibinfo{volume}{22}(\bibinfo{number}{1}), pp. \bibinfo{pages}{11--32},
  \doi{10.3166/tsi.22.11-32}.

\bibitemdeclare{article}{carla}
\bibitem{carla}
\bibinfo{author}{Alexey \surnamestart Dosovitskiy\surnameend},
  \bibinfo{author}{Germ{\'{a}}n \surnamestart Ros\surnameend},
  \bibinfo{author}{Felipe \surnamestart Codevilla\surnameend},
  \bibinfo{author}{Antonio~M. \surnamestart L{\'{o}}pez\surnameend} \&
  \bibinfo{author}{Vladlen \surnamestart Koltun\surnameend}
  (\bibinfo{year}{2017}): \emph{\bibinfo{title}{{CARLA:} An Open Urban Driving
  Simulator}}.
\newblock {\slshape \bibinfo{journal}{CoRR}} \bibinfo{volume}{abs/1711.03938},
  \doi{10.48550/arXiv.1711.03938}.
\newblock \eprint{1711.03938}.

\bibitemdeclare{article}{trainsimdataset}
\bibitem{trainsimdataset}
\bibinfo{author}{Gianluca \surnamestart D’Amico\surnameend},
  \bibinfo{author}{Mauro \surnamestart Marinoni\surnameend},
  \bibinfo{author}{Federico \surnamestart Nesti\surnameend},
  \bibinfo{author}{Giulio \surnamestart Rossolini\surnameend},
  \bibinfo{author}{Giorgio \surnamestart Buttazzo\surnameend},
  \bibinfo{author}{Salvatore \surnamestart Sabina\surnameend} \&
  \bibinfo{author}{Gianluigi \surnamestart Lauro\surnameend}
  (\bibinfo{year}{2023}): \emph{\bibinfo{title}{TrainSim: A Railway Simulation
  Framework for LiDAR and Camera Dataset Generation}}.
\newblock {\slshape \bibinfo{journal}{IEEE Transactions on Intelligent
  Transportation Systems}} \bibinfo{volume}{24}(\bibinfo{number}{12}), pp.
  \bibinfo{pages}{15006--15017}, \doi{10.1109/TITS.2023.3297728}.

\bibitemdeclare{inproceedings}{Siemens:B2007}
\bibitem{Siemens:B2007}
\bibinfo{author}{Didier \surnamestart Essam\'{e}\surnameend} \&
  \bibinfo{author}{Daniel \surnamestart Doll\'{e}\surnameend}
  (\bibinfo{year}{2007}): \emph{\bibinfo{title}{B in Large Scale Projects: The
  {C}anarsie Line {CBTC} Experience}}.
\newblock In: {\slshape \bibinfo{booktitle}{Proceedings {B} (B2007)}},
  \bibinfo{series}{LNCS 4355}, \bibinfo{publisher}{Springer},
  \bibinfo{address}{Besancon, France}, pp. \bibinfo{pages}{252--254},
  \doi{10.1007/11955757_21}.

\bibitemdeclare{inproceedings}{gehr2018ai2}
\bibitem{gehr2018ai2}
\bibinfo{author}{Timon \surnamestart Gehr\surnameend}, \bibinfo{author}{Matthew
  \surnamestart Mirman\surnameend}, \bibinfo{author}{Dana \surnamestart
  Drachsler-Cohen\surnameend}, \bibinfo{author}{Petar \surnamestart
  Tsankov\surnameend}, \bibinfo{author}{Swarat \surnamestart
  Chaudhuri\surnameend} \& \bibinfo{author}{Martin \surnamestart
  Vechev\surnameend} (\bibinfo{year}{2018}): \emph{\bibinfo{title}{Ai2: Safety
  and robustness certification of neural networks with abstract
  interpretation}}.
\newblock In: {\slshape \bibinfo{booktitle}{2018 IEEE symposium on security and
  privacy (SP)}}, \bibinfo{organization}{IEEE}, pp. \bibinfo{pages}{3--18},
  \doi{10.1109/SP.2018.00058}.

\bibitemdeclare{inproceedings}{robustness}
\bibitem{robustness}
\bibinfo{author}{Divya \surnamestart Gopinath\surnameend}, \bibinfo{author}{Guy
  \surnamestart Katz\surnameend}, \bibinfo{author}{Corina~S. \surnamestart
  P{\u{a}}s{\u{a}}reanu\surnameend} \& \bibinfo{author}{Clark \surnamestart
  Barrett\surnameend} (\bibinfo{year}{2018}): \emph{\bibinfo{title}{DeepSafe: A
  Data-Driven Approach for Assessing Robustness of Neural Networks}}.
\newblock In: {\slshape \bibinfo{booktitle}{Proceedings ATVA}},
  \bibinfo{series}{LNCS 11138}, \bibinfo{publisher}{Springer}, pp.
  \bibinfo{pages}{3--19}, \doi{10.1007/978-3-030-01090-4_1}.

\bibitemdeclare{inproceedings}{fraunhoferSimEnv}
\bibitem{fraunhoferSimEnv}
\bibinfo{author}{J{\"u}rgen \surnamestart Grossmann\surnameend},
  \bibinfo{author}{Nicolas \surnamestart Grube\surnameend},
  \bibinfo{author}{Sami \surnamestart Kharma\surnameend},
  \bibinfo{author}{Dorian \surnamestart Knoblauch\surnameend},
  \bibinfo{author}{Roman \surnamestart Krajewski\surnameend},
  \bibinfo{author}{Mariia \surnamestart Kucheiko\surnameend} \&
  \bibinfo{author}{Hans-Werner \surnamestart Wiesbrock\surnameend}
  (\bibinfo{year}{2023}): \emph{\bibinfo{title}{Test and Training Data
  Generation for Object Recognition in the Railway Domain}}.
\newblock In: {\slshape \bibinfo{booktitle}{SEFM 2022 Collocated Workshops}},
  {\slshape \bibinfo{series}{LNCS}} \bibinfo{volume}{13765},
  \bibinfo{organization}{Springer}, pp. \bibinfo{pages}{5--16},
  \doi{10.1007/978-3-031-26236-4_1}.

\bibitemdeclare{inproceedings}{gruteser2023formal}
\bibitem{gruteser2023formal}
\bibinfo{author}{Jan \surnamestart Gruteser\surnameend}, \bibinfo{author}{David
  \surnamestart Gele{\ss}us\surnameend}, \bibinfo{author}{Michael \surnamestart
  Leuschel\surnameend}, \bibinfo{author}{Jan \surnamestart
  Ro{\ss}bach\surnameend} \& \bibinfo{author}{Fabian \surnamestart
  Vu\surnameend} (\bibinfo{year}{2023}): \emph{\bibinfo{title}{A Formal Model
  of Train Control with AI-based Obstacle Detection}}.
\newblock In: {\slshape \bibinfo{booktitle}{Proceedings RSSRail}}, {\slshape
  \bibinfo{series}{LNCS}} \bibinfo{volume}{14198},
  \bibinfo{organization}{Springer}, pp. \bibinfo{pages}{128--145},
  \doi{10.1007/978-3-031-43366-5_8}.

\bibitemdeclare{inproceedings}{gruteser-railml}
\bibitem{gruteser-railml}
\bibinfo{author}{Jan \surnamestart Gruteser\surnameend} \&
  \bibinfo{author}{Michael \surnamestart Leuschel\surnameend}
  (\bibinfo{year}{2024}): \emph{\bibinfo{title}{Validation of RailML Using
  ProB}}.
\newblock In: {\slshape \bibinfo{booktitle}{Proceedings ICECCS 2024}},
  {\slshape \bibinfo{series}{LNCS}} \bibinfo{volume}{14784},
  \bibinfo{organization}{Springer}, pp. \bibinfo{pages}{245--256},
  \doi{10.1007/978-3-031-66456-4_13}.

\bibitemdeclare{article}{DBLP:journals/sttt/HansenLKKNNSS20}
\bibitem{DBLP:journals/sttt/HansenLKKNNSS20}
\bibinfo{author}{Dominik \surnamestart Hansen\surnameend},
  \bibinfo{author}{Michael \surnamestart Leuschel\surnameend},
  \bibinfo{author}{Philipp \surnamestart K{\"{o}}rner\surnameend},
  \bibinfo{author}{Sebastian \surnamestart Krings\surnameend},
  \bibinfo{author}{Thomas \surnamestart Naulin\surnameend},
  \bibinfo{author}{Nader \surnamestart Nayeri\surnameend},
  \bibinfo{author}{David \surnamestart Schneider\surnameend} \&
  \bibinfo{author}{Frank \surnamestart Skowron\surnameend}
  (\bibinfo{year}{2020}): \emph{\bibinfo{title}{Validation and real-life
  demonstration of {ETCS} hybrid level 3 principles using a formal {B} model}}.
\newblock {\slshape \bibinfo{journal}{Int. J. Softw. Tools Technol. Transf.}}
  \bibinfo{volume}{22}(\bibinfo{number}{3}), pp. \bibinfo{pages}{315--332},
  \doi{10.1007/s10009-020-00551-6}.

\bibitemdeclare{article}{hensel2022probabilistic}
\bibitem{hensel2022probabilistic}
\bibinfo{author}{Christian \surnamestart Hensel\surnameend},
  \bibinfo{author}{Sebastian \surnamestart Junges\surnameend},
  \bibinfo{author}{Joost-Pieter \surnamestart Katoen\surnameend},
  \bibinfo{author}{Tim \surnamestart Quatmann\surnameend} \&
  \bibinfo{author}{Matthias \surnamestart Volk\surnameend}
  (\bibinfo{year}{2022}): \emph{\bibinfo{title}{{The probabilistic model
  checker Storm}}}.
\newblock {\slshape \bibinfo{journal}{STTT}}
  \bibinfo{volume}{24}(\bibinfo{number}{4}), pp. \bibinfo{pages}{589--610},
  \doi{10.1007/s10009-021-00633-z}.

\bibitemdeclare{inproceedings}{verification-dnn}
\bibitem{verification-dnn}
\bibinfo{author}{Xiaowei \surnamestart Huang\surnameend},
  \bibinfo{author}{Marta \surnamestart Kwiatkowska\surnameend},
  \bibinfo{author}{Sen \surnamestart Wang\surnameend} \& \bibinfo{author}{Min
  \surnamestart Wu\surnameend} (\bibinfo{year}{2017}):
  \emph{\bibinfo{title}{Safety Verification of Deep Neural Networks}}.
\newblock In: {\slshape \bibinfo{booktitle}{Proceedings CAV}},
  \bibinfo{series}{LNCS 10426}, \bibinfo{publisher}{Springer}, pp.
  \bibinfo{pages}{3--29}, \doi{10.1007/978-3-319-63387-9_1}.

\bibitemdeclare{article}{certcontrol}
\bibitem{certcontrol}
\bibinfo{author}{Daniel \surnamestart Jackson\surnameend},
  \bibinfo{author}{Valerie \surnamestart Richmond\surnameend},
  \bibinfo{author}{Mike \surnamestart Wang\surnameend}, \bibinfo{author}{Jeff
  \surnamestart Chow\surnameend}, \bibinfo{author}{Uriel \surnamestart
  Guajardo\surnameend}, \bibinfo{author}{Soonho \surnamestart Kong\surnameend},
  \bibinfo{author}{Sergio \surnamestart Campos\surnameend},
  \bibinfo{author}{Geoffrey \surnamestart Litt\surnameend} \&
  \bibinfo{author}{Nikos \surnamestart Ar{\'{e}}chiga\surnameend}
  (\bibinfo{year}{2021}): \emph{\bibinfo{title}{Certified Control: An
  Architecture for Verifiable Safety of Autonomous Vehicles}}.
\newblock {\slshape \bibinfo{journal}{CoRR}} \bibinfo{volume}{abs/2104.06178},
  \doi{10.48550/arXiv.2104.06178}.
\newblock \eprint{2104.06178}.

\bibitemdeclare{inproceedings}{taxinet}
\bibitem{taxinet}
\bibinfo{author}{Ismet~Burak \surnamestart Kadron\surnameend},
  \bibinfo{author}{Divya \surnamestart Gopinath\surnameend},
  \bibinfo{author}{Corina~S. \surnamestart P{\u{a}}s{\u{a}}reanu\surnameend} \&
  \bibinfo{author}{Huafeng \surnamestart Yu\surnameend} (\bibinfo{year}{2022}):
  \emph{\bibinfo{title}{Case Study: Analysis of Autonomous Center Line Tracking
  Neural Networks}}.
\newblock In: {\slshape \bibinfo{booktitle}{Proceedings VSTTE 2021}},
  \bibinfo{series}{LNCS 13124}, \bibinfo{publisher}{Springer}, pp.
  \bibinfo{pages}{104--121}, \doi{10.1007/978-3-030-95561-8_7}.

\bibitemdeclare{inproceedings}{katz2017reluplex}
\bibitem{katz2017reluplex}
\bibinfo{author}{Guy \surnamestart Katz\surnameend}, \bibinfo{author}{Clark
  \surnamestart Barrett\surnameend}, \bibinfo{author}{David~L \surnamestart
  Dill\surnameend}, \bibinfo{author}{Kyle \surnamestart Julian\surnameend} \&
  \bibinfo{author}{Mykel~J \surnamestart Kochenderfer\surnameend}
  (\bibinfo{year}{2017}): \emph{\bibinfo{title}{Reluplex: An efficient SMT
  solver for verifying deep neural networks}}.
\newblock In: {\slshape \bibinfo{booktitle}{Proceedings CAV}},
  \bibinfo{series}{LNCS 10426}, \bibinfo{publisher}{Springer}, pp.
  \bibinfo{pages}{97--117}, \doi{10.1007/978-3-319-63387-9_5}.

\bibitemdeclare{inproceedings}{kwiatkowska2011prism}
\bibitem{kwiatkowska2011prism}
\bibinfo{author}{Marta \surnamestart Kwiatkowska\surnameend},
  \bibinfo{author}{Gethin \surnamestart Norman\surnameend} \&
  \bibinfo{author}{David \surnamestart Parker\surnameend}
  (\bibinfo{year}{2011}): \emph{\bibinfo{title}{{PRISM 4.0: Verification of
  probabilistic real-time systems}}}.
\newblock In: {\slshape \bibinfo{booktitle}{Proceedings CAV}},
  \bibinfo{volume}{LNCS 6806}, \bibinfo{publisher}{Springer}, pp.
  \bibinfo{pages}{585--591}, \doi{10.1007/978-3-642-22110-1_47}.

\bibitemdeclare{article}{probjournal}
\bibitem{probjournal}
\bibinfo{author}{Michael \surnamestart Leuschel\surnameend} \&
  \bibinfo{author}{Michael \surnamestart Butler\surnameend}
  (\bibinfo{year}{2008}): \emph{\bibinfo{title}{{ProB}: an automated analysis
  toolset for the {B} method}}.
\newblock {\slshape \bibinfo{journal}{STTT}}
  \bibinfo{volume}{10}(\bibinfo{number}{2}), pp. \bibinfo{pages}{185--203},
  \doi{10.1007/s10009-007-0063-9}.

\bibitemdeclare{inproceedings}{etcs-proof}
\bibitem{etcs-proof}
\bibinfo{author}{Michael \surnamestart Leuschel\surnameend} \&
  \bibinfo{author}{Nader \surnamestart Nayeri\surnameend}
  (\bibinfo{year}{2023}): \emph{\bibinfo{title}{Modelling, Visualisation and
  Proof of an ETCS Level 3 Moving Block System}}.
\newblock In: {\slshape \bibinfo{booktitle}{Proceedings RSSRail}},
  \bibinfo{series}{LNCS 14198}, \bibinfo{publisher}{Springer}, pp.
  \bibinfo{pages}{193--210}, \doi{10.1007/978-3-031-43366-5_12}.

\bibitemdeclare{article}{literatureValidationAiSystems}
\bibitem{literatureValidationAiSystems}
\bibinfo{author}{Lalli \surnamestart Myllyaho\surnameend},
  \bibinfo{author}{Mikko \surnamestart Raatikainen\surnameend},
  \bibinfo{author}{Tomi \surnamestart Männistö\surnameend},
  \bibinfo{author}{Tommi \surnamestart Mikkonen\surnameend} \&
  \bibinfo{author}{Jukka~K. \surnamestart Nurminen\surnameend}
  (\bibinfo{year}{2021}): \emph{\bibinfo{title}{Systematic literature review of
  validation methods for AI systems}}.
\newblock {\slshape \bibinfo{journal}{Journal of Systems and Software}}
  \bibinfo{volume}{181}, p. \bibinfo{pages}{111050},
  \doi{10.1016/j.jss.2021.111050}.

\bibitemdeclare{book}{nonami2010autonomous}
\bibitem{nonami2010autonomous}
\bibinfo{author}{Kenzo \surnamestart Nonami\surnameend}, \bibinfo{author}{Farid
  \surnamestart Kendoul\surnameend}, \bibinfo{author}{Satoshi \surnamestart
  Suzuki\surnameend}, \bibinfo{author}{Wei \surnamestart Wang\surnameend} \&
  \bibinfo{author}{Daisuke \surnamestart Nakazawa\surnameend}
  (\bibinfo{year}{2010}): \emph{\bibinfo{title}{Autonomous flying robots:
  unmanned aerial vehicles and micro aerial vehicles}}.
\newblock \bibinfo{publisher}{Springer Science \& Business Media},
  \doi{10.1007/978-4-431-53856-1}.

\bibitemdeclare{inproceedings}{perception-DNN}
\bibitem{perception-DNN}
\bibinfo{author}{Corina~S. \surnamestart P{\u{a}}s{\u{a}}reanu\surnameend},
  \bibinfo{author}{Ravi \surnamestart Mangal\surnameend},
  \bibinfo{author}{Divya \surnamestart Gopinath\surnameend},
  \bibinfo{author}{Sinem \surnamestart Getir~Yaman\surnameend},
  \bibinfo{author}{Calum \surnamestart Imrie\surnameend}, \bibinfo{author}{Radu
  \surnamestart Calinescu\surnameend} \& \bibinfo{author}{Huafeng \surnamestart
  Yu\surnameend} (\bibinfo{year}{2023}): \emph{\bibinfo{title}{Closed-Loop
  Analysis of Vision-Based Autonomous Systems: A Case Study}}.
\newblock In: {\slshape \bibinfo{booktitle}{Proceedings CAV}},
  \bibinfo{series}{LNCS 13964}, \bibinfo{publisher}{Springer}, pp.
  \bibinfo{pages}{289--303}, \doi{10.1007/978-3-031-37706-8_15}.

\bibitemdeclare{inproceedings}{peleska2022standardisation}
\bibitem{peleska2022standardisation}
\bibinfo{author}{Jan \surnamestart Peleska\surnameend}, \bibinfo{author}{Anne~E
  \surnamestart Haxthausen\surnameend} \& \bibinfo{author}{Thierry
  \surnamestart Lecomte\surnameend} (\bibinfo{year}{2022}):
  \emph{\bibinfo{title}{Standardisation considerations for autonomous train
  control}}.
\newblock In: {\slshape \bibinfo{booktitle}{Proceedings ISoLA}}, {\slshape
  \bibinfo{series}{LNCS}} \bibinfo{volume}{13704},
  \bibinfo{organization}{Springer}, pp. \bibinfo{pages}{286--307},
  \doi{10.1007/978-3-031-19762-8_22}.

\bibitemdeclare{inproceedings}{DBLP:journals/corr/RedmonDGF15}
\bibitem{DBLP:journals/corr/RedmonDGF15}
\bibinfo{author}{Joseph \surnamestart Redmon\surnameend},
  \bibinfo{author}{Santosh~Kumar \surnamestart Divvala\surnameend},
  \bibinfo{author}{Ross~B. \surnamestart Girshick\surnameend} \&
  \bibinfo{author}{Ali \surnamestart Farhadi\surnameend}
  (\bibinfo{year}{2016}): \emph{\bibinfo{title}{{You Only Look Once: Unified,
  Real-Time Object Detection}}}.
\newblock In: {\slshape \bibinfo{booktitle}{2016 IEEE Conference on Computer
  Vision and Pattern Recognition (CVPR)}}, \bibinfo{publisher}{IEEE Computer
  Society}, \bibinfo{address}{Los Alamitos, CA, USA}, pp.
  \bibinfo{pages}{779--788}, \doi{10.1109/CVPR.2016.91}.

\bibitemdeclare{inproceedings}{Rossbach2024}
\bibitem{Rossbach2024}
\bibinfo{author}{Jan \surnamestart Roßbach\surnameend},
  \bibinfo{author}{Oliver~De \surnamestart Candido\surnameend},
  \bibinfo{author}{Ahmed \surnamestart Hamman\surnameend} \&
  \bibinfo{author}{Michael \surnamestart Leuschel\surnameend}
  (\bibinfo{year}{2024}): \emph{\bibinfo{title}{Evaluating AI-based Components
  for Autonomous Railway System}}.
\newblock In: {\slshape \bibinfo{booktitle}{Proceedings KI 2024}}, {\slshape
  \bibinfo{series}{LNAI}} \bibinfo{volume}{14992},
  \bibinfo{publisher}{Springer}, pp. \bibinfo{pages}{190--203},
  \doi{10.1007/978-3-031-70893-0_14}.

\bibitemdeclare{article}{Rossbach2023}
\bibitem{Rossbach2023}
\bibinfo{author}{Jan \surnamestart Roßbach\surnameend} \&
  \bibinfo{author}{Michael \surnamestart Leuschel\surnameend}
  (\bibinfo{year}{2023}): \emph{\bibinfo{title}{Certified Control for Train
  Sign Classification}}.
\newblock {\slshape \bibinfo{journal}{EPTCS}} \bibinfo{volume}{395}, pp.
  \bibinfo{pages}{69--76}, \doi{10.4204/eptcs.395.5}.

\bibitemdeclare{inproceedings}{ruan2018reachability}
\bibitem{ruan2018reachability}
\bibinfo{author}{Wenjie \surnamestart Ruan\surnameend},
  \bibinfo{author}{Xiaowei \surnamestart Huang\surnameend} \&
  \bibinfo{author}{Marta \surnamestart Kwiatkowska\surnameend}
  (\bibinfo{year}{2018}): \emph{\bibinfo{title}{{Reachability Analysis of Deep
  Neural Networks with Provable Guarantees}}}.
\newblock In: {\slshape \bibinfo{booktitle}{Proceedings IJCAI}}, pp.
  \bibinfo{pages}{2651--2659}, \doi{10.24963/ijcai.2018/368}.

\bibitemdeclare{inproceedings}{sun2020scalability}
\bibitem{sun2020scalability}
\bibinfo{author}{Pei \surnamestart Sun\surnameend}, \bibinfo{author}{Henrik
  \surnamestart Kretzschmar\surnameend}, \bibinfo{author}{Xerxes \surnamestart
  Dotiwalla\surnameend}, \bibinfo{author}{Aurelien \surnamestart
  Chouard\surnameend}, \bibinfo{author}{Vijaysai \surnamestart
  Patnaik\surnameend}, \bibinfo{author}{Paul \surnamestart Tsui\surnameend},
  \bibinfo{author}{James \surnamestart Guo\surnameend}, \bibinfo{author}{Yin
  \surnamestart Zhou\surnameend}, \bibinfo{author}{Yuning \surnamestart
  Chai\surnameend}, \bibinfo{author}{Benjamin \surnamestart Caine\surnameend}
  et~al. (\bibinfo{year}{2020}): \emph{\bibinfo{title}{Scalability in
  perception for autonomous driving: Waymo open dataset}}.
\newblock In: {\slshape \bibinfo{booktitle}{Proceedings of the IEEE/CVF
  conference on computer vision and pattern recognition}}, pp.
  \bibinfo{pages}{2446--2454}, \doi{10.48550/arXiv.1912.04838}.

\bibitemdeclare{inproceedings}{vu2024validation}
\bibitem{vu2024validation}
\bibinfo{author}{Fabian \surnamestart Vu\surnameend}, \bibinfo{author}{Jannik
  \surnamestart Dunkelau\surnameend} \& \bibinfo{author}{Michael \surnamestart
  Leuschel\surnameend} (\bibinfo{year}{2024}): \emph{\bibinfo{title}{Validation
  of Reinforcement Learning Agents and Safety Shields with ProB}}.
\newblock In: {\slshape \bibinfo{booktitle}{NASA Formal Methods Symposium}},
  {\slshape \bibinfo{series}{LNCS}} \bibinfo{volume}{14627},
  \bibinfo{organization}{Springer}, pp. \bibinfo{pages}{279--297},
  \doi{10.1007/978-3-031-60698-4_16}.

\bibitemdeclare{inproceedings}{simb}
\bibitem{simb}
\bibinfo{author}{Fabian \surnamestart Vu\surnameend}, \bibinfo{author}{Michael
  \surnamestart Leuschel\surnameend} \& \bibinfo{author}{Atif \surnamestart
  Mashkoor\surnameend} (\bibinfo{year}{2021}):
  \emph{\bibinfo{title}{{Validation of Formal Models by Timed Probabilistic
  Simulation}}}.
\newblock In: {\slshape \bibinfo{booktitle}{Proceedings ABZ}}, {\slshape
  \bibinfo{series}{LNCS}} \bibinfo{volume}{12709},
  \bibinfo{publisher}{Springer}, pp. \bibinfo{pages}{81--96},
  \doi{10.1007/978-3-030-77543-8_6}.

\bibitemdeclare{inproceedings}{visb}
\bibitem{visb}
\bibinfo{author}{Michelle \surnamestart Werth\surnameend} \&
  \bibinfo{author}{Michael \surnamestart Leuschel\surnameend}
  (\bibinfo{year}{2020}): \emph{\bibinfo{title}{{VisB}: A Lightweight Tool to
  Visualize Formal Models with {SVG} Graphics}}.
\newblock In: {\slshape \bibinfo{booktitle}{Proceedings ABZ}},
  \bibinfo{series}{LNCS 12071}, \bibinfo{publisher}{Springer}, pp.
  \bibinfo{pages}{260--265}, \doi{10.1007/978-3-030-48077-6_21}.

\bibitemdeclare{inproceedings}{wild2023towards}
\bibitem{wild2023towards}
\bibinfo{author}{Michael \surnamestart Wild\surnameend},
  \bibinfo{author}{Jan~Steffen \surnamestart Becker\surnameend},
  \bibinfo{author}{G{\"u}nter \surnamestart Ehmen\surnameend} \&
  \bibinfo{author}{Eike \surnamestart M{\"o}hlmann\surnameend}
  (\bibinfo{year}{2023}): \emph{\bibinfo{title}{Towards Scenario-Based
  Certification of Highly Automated Railway Systems}}.
\newblock In: {\slshape \bibinfo{booktitle}{Proceedings RSSRail}}, {\slshape
  \bibinfo{series}{LNCS}} \bibinfo{volume}{14198},
  \bibinfo{organization}{Springer}, pp. \bibinfo{pages}{78--97},
  \doi{10.1007/978-3-031-43366-5_5}.

\end{thebibliography}
\end{document}